\documentclass[twocolumn]{aastex62}

\usepackage{epsfig}
\usepackage{epstopdf}
\usepackage{graphics}
\usepackage{amsmath}
\usepackage{xcolor}
\usepackage{aas_macros}

\usepackage{array}
\usepackage{booktabs}
\usepackage{amsmath,amssymb,amsfonts,amsbsy}
\usepackage{mathrsfs}
\usepackage{url}
\usepackage{multirow}
\usepackage{xcolor}
\usepackage{color}
\usepackage{ulem}
\usepackage{enumerate}

\newcommand{\dif}{\mathrm{d}}

\renewcommand{\figurename}{Fig.}
\renewcommand{\tablename}{Tab.}
\begin{document}

\title{Galactic cosmic ray propagation: sub-PeV diffuse gamma-ray and neutrino emission}

\correspondingauthor{Wei Liu, Meng-Jie Zhao}
\email{liuwei@ihep.ac.cn, zhaomj@ihep.ac.cn}

\author{Bing-Qiang Qiao}
\affiliation{Key Laboratory of Particle Astrophysics,
	Institute of High Energy Physics, Chinese Academy of Sciences, Beijing 100049, China
}

\author{Wei Liu}
\affiliation{Key Laboratory of Particle Astrophysics,
	Institute of High Energy Physics, Chinese Academy of Sciences, Beijing 100049, China
}

\author{Meng-Jie Zhao}
\affiliation{Key Laboratory of Particle Astrophysics,
	Institute of High Energy Physics, Chinese Academy of Sciences, Beijing 100049, China
}
\affiliation{University of Chinese Academy of Sciences, Beijing 100049, China
}

\author{Xiao-Jun Bi}
\affiliation{Key Laboratory of Particle Astrophysics,
	Institute of High Energy Physics, Chinese Academy of Sciences, Beijing 100049, China
}
\affiliation{University of Chinese Academy of Sciences, Beijing 100049, China
}

\author{Yi-Qing Guo}
\affiliation{Key Laboratory of Particle Astrophysics,
	Institute of High Energy Physics, Chinese Academy of Sciences, Beijing 100049, China
}



\begin{abstract}
The Tibet AS$\gamma$ experiment just reported their measurement of sub-PeV diffuse gamma ray emission from the Galactic disk, with the highest energy up to $957$ TeV. These gamma-rays are most likely the hadronic origin by cosmic ray interaction with interstellar gas in the Galaxy. This measurement provides direct evidence to the hypothesis that the Galactic cosmic rays can be accelerated beyond PeV energies. In this work, we try to explain the sub-PeV diffuse gamma-ray spectrum within cosmic rays diffusive propagation model. We find there is a tension between the sub-PeV diffuse gamma rays and the local cosmic ray spectrum. To describe the sub-PeV diffuse gamma-ray flux, it generally requires larger local cosmic-ray flux than measurement in the knee region.
We further calculate the PeV neutrino flux from the cosmic ray propagation model. Even all of these sub-PeV diffuse gamma rays originate from the propagation, the Galactic neutrinos only account for less than $\sim 15\%$ of observed flux, most of which are still from extragalactic sources.
\end{abstract}


\section{Introduction}

In the all-particle spectrum, the most prominent feature is the so-called ``knee" structure at $\sim 4$ PeV, where the spectrum exhibits a slight steepening, with the slope changing from $-2.7$ to $-3.1$ \citep{2008ApJ...678.1165A}. Meanwhile above the knee, a tendency to the heavier nuclei has been noted \citep{2012APh....35..660K, 2018mmea.book....1A}. It is believed that the cosmic rays (CRs) less than PeV energies are principally of Galactic origin. However, which Galactic objects could accelerate PeV CRs, have long been in dispute. The supernova remnants (SNRs) had long been proposed as the sources of Galactic CRs \citep{1934CoMtW...3...79B}. But long gazes into most SNRs associated with the molecular clouds indicate that the cut-off energy in the gamma-ray spectrum is less than $10$ TeV, which means the inferred maximum CR energy is below $\sim 100$ TeV.

Massive strides about the PeV sources have been made in recent years. The Galactic center, which embodies a supermassive black holes, Sagittarius A$^\ast$, with mass up to $4 \times 10^6$ solar mass, is regarded as a PeVatron. The observations of anisotropy phase also seems to hint at this \citep{2012ApJ...746...33A, 2016ApJ...826..220A, 2017ApJ...836..153A}. The HESS collaboration performed deep observations of the central molecular zone surrounding the Galactic center. They detected a single power-law gamma-ray spectrum up to tens of TeV, without a break or a cutoff. This strongly indicates a PeV proton accelerator at the Galactic center \citep{2016Natur.531..476H}. Early in this year, the Tibet AS$\gamma$ experiment reported their observation of SNR G106.3+2.7 \citep{2021NatAs.tmp...41T}. They successfully detected the $100$ TeV gamma-ray photons, which are well correlated with the nearby molecular cloud. Therefore the morphological feature favours a hadronic origin of such very-high energy gamma rays. The young massive star clusters are another possible acceleration sites. Most recently, HAWC published their gamma-ray observations of the Cyngnus Cocoon, a well-known superbubble, with energy ranging from $1$ to $100$ TeV \citep{2021arXiv210306820A}. The $100$ TeV gamma-ray photons may originate from the accelerated CRs in the Cyg OB2, an active star-forming region. These recent discoveries provide hard evidence concerning the Galactic origin of PeV CRs.

Apart from searching for point sources, the observations of $100$ TeV diffuse gamma ray emission could also be served as the evidence of PeV CR origin. At such high energy, gamma ray photons are expected to have interactions with the low energy background radiation, so that the extragalactic flux is strongly suppressed. On the other hand, the all-electron spectrum shows a clear steepening above $1$ TeV, where the spectral index changes to $\sim -3.9$ \citep{ 2017Natur.552...63D, hess-icrc2017}. Coupled with the energy loss during propagation of electrons, the $100$ TeV gamma-ray flux from the inverse Compton scattering off electrons could be safely neglected. Such energetic gamma rays are clearly generated from the $\pi^0$-decay process during the hadronic interaction of the CR nuclei with the interstellar medium (ISM). Previously, only CASA-MIA \citep{1998ApJ...493..175B} and KASCADE \citep{2017ApJ...848....1A} experiments set upper limits on the diffuse gamma rays above $20$ TeV. At present, the Tibet AS$\gamma$ experiment managed to detect the sub-PeV diffuse gamma rays in the Galactic disk, with energy ranging from $100$ TeV to $1$ PeV \citep{PhysRevLett.126.141101}. The observed highest energy is up to $957$ TeV, very close to $1$ PeV. This discovery indicates that the Galactic sources could accelerate CRs to at least  $\sim 10$ PeV.

During the p-p interactions of CRs with ISM, the generated gamma-ray photons are simultaneously accompanied with the neutrinos, which are the decay products of the $\pi^{\pm}$s. Thereupon, the high-energy neutrinos are also regarded as a good probe to the hadronic interaction. The astrophysical neutrinos have been detected by the IceCube Neutrino Observatory \citep{2013Sci...342E...1I, 2013PhRvL.111b1103A, 2014PhRvL.113j1101A}. The overall distribution of the neutrino event samples is consistent with an isotropic distribution, which demonstrates that they are held to be predominantly extragalactic. The directional searches also have found an excess from a starburst galaxy \citep{2020PhRvL.124e1103A} and the neutrino emission associated with a blazar \citep{2018Sci...361.1378I, 2018Sci...361..147I}. However the proportion of Galactic contribution is still uncertain. A separate fit of the Northern and Southern hemisphere signals in the four-year signal shows a preference to a harder spectrum in the Northern hemisphere \citep{2015ApJ...809...98A} which could potentially be due to the presence of a softer contribution of the flux from the inner Galaxy in the Southern hemisphere.


The sub-PeV diffuse gamma-ray emission effectively traces the spatial distribution of remote CRs, so they could be well applied to testify the available propagation models. Besides, it is helpful to unveil the origin of knee region \citep{2014ApJ...795..100G}. In this work, we investigate the propagation origin of sub-PeV diffuse gamma rays. We find there is a tension between sub-PeV diffuse gamma rays and local cosmic ray measurements. To explain the sub-PeV diffuse gamma-rays measured by the Tibet experiment, the calculated local cosmic-ray flux inevitably exceeds the local CR flux above PeV energies. One possibility is that the propagated spectrum is harder close to the Galactic center as for the current propagation models, whereas the propagation around the solar system is still unchanged. Meanwhile these sub-PeV gamma-ray photons from point sources, especially unresolved, may not entirely subtracted. We further evaluate the Galactic contribution to the diffuse neutrino flux and find at most $15 \%$ of observed flux come from Galactic CR propagation. \cite{2018PhRvD..98d3003L} has made the predictions for the diffuse gamm-rays above hundreds of TeV. In their work, the spatial distribution of CRs is extrapolated according to the local CR measurement and diffuse gamma observations by Fermi-LAT. In this work, we derive the CR spatial distribution by solving the diffusive propagation model.


\section{Model Description}

\subsection{Homogeneous diffusion}
In the conventional propagation model, the diffusion process is supposed to be homogeneous and isotropic, so the diffusion coefficient is only a function of rigidity ${\cal R} = p/Ze$, namely,
\begin{equation}
D({\cal R}) = D_0 \beta^\eta \left( \dfrac{\cal R}{ {\cal R}_0 }\right)^{\delta_0} ~.
\end{equation}
The power index $\delta$ is usually taken from $0.3$ to $0.6$, as inferred from the fitting of boron-to-carbon ratio \citep{2017PhRvD..95h3007Y}. After propagation, the CR spectrum falls off as a single power-law, $\phi \propto {\cal R}^{-\nu -\delta}$, where $\nu$ is the power index of CR spectrum at source.

\subsection{Spatial-dependent propagation}
However the spectral hardening of CR nuclei above $\sim 200$ GV \citep{2011Sci...332...69A} as well as the anisotropy observations \citep{2017ApJ...836..153A} severely challenge the conventional homogeneous propagation model \citep{2012JCAP...01..011B, 2017PhRvD..96b3006L}. The spatial-dependent propagation (SDP) was intially introduced to account for the excess of CR nuclei \citep{2012ApJ...752L..13T}. Afterwards, it is further applied to large-scale anisotropy \citep{2019JCAP...10..010L, 2019JCAP...12..007Q} and diffuse gamma-rays \citep{2018PhRvD..97f3008G} observations. For a comprehensive introduction, one can refer to \cite{2016ApJ...819...54G} and \cite{2018ApJ...869..176L}.

In contrast to the homogeneous diffusion, the entire diffusive halo in the SDP model is split into two zones characterized by diverse diffusion properties, i.e. inner halo (IH) and outer halo (OH). The Galactic disk and its surrounding areas within a few hundred parsecs are called IH, where the diffusion is slower and relevant to the radial distribution of sources. The diffusion in extended regions outside of IH, i.e. OH, is faster and approaches to the conventional propagation. The diffusion coefficient $D$ in the whole region is thus parameterized as \citep{2016ApJ...819...54G, 2018ApJ...869..176L}:
\begin{equation}
D(r,z, {\cal R} )= D_{0}F(r,z)\beta^{\eta} \left(\dfrac{\cal R}{{\cal R}_{0}} \right)^{\delta_0 F(r,z)} ~.
\label{eq:diffusion}
\end{equation}
with
\begin{equation}
F(r,z) =
\begin{cases}
g(r,z) +\left[1-g(r,z) \right] \left(\dfrac{z}{\xi L} \right)^{n} , &  |z| \leq \xi L \\
1 ~, & |z| > \xi L
\end{cases} ~.
\end{equation}
$g(r,z)$ is $N_m/[1+f(r,z)]$, in which $f(r,z)$ is the source density distribution.

\section{Results}

In this work, to make a detailed study of the propagation origin of sub-PeV diffuse gamma-rays, we compare two kinds of common propagation scenarios, i.e. homogeneous diffusion (HD) and spatial-dependent propagation (SDP). As for the SDP model, another two models are introduced, according to the origin of spectral harding. In the model SDP-A, the excesses of CR nuclei above $200$ GV are regarded as a local effect, which originate from a local SNR. A nearby SNR is also favored in order to describe the evolution of anisotropy amplitude and phase with energy \citep{2016PhRvL.117o1103A, 2019JCAP...10..010L, 2019JCAP...12..007Q}. For model SDP-B, the excesses chiefly originate from the spatial variation of diffusion coefficient. The excess of diffuse gamma-ray emission at Galactic plane above a few GeV has also been well account for in model SDP-B \citep{2018PhRvD..97f3008G}, due to the spectral hardening of CRs in the whole diffusive halo. In the calculations below, the injection spectra of all propagation models are assumed to have a power-law plus a high-energy exponential cutoff, i.e.
\begin{equation}
q_i ({\cal R})=q^i_{0}
\left(\dfrac{{\cal R}}{{\cal R}_{\rm 0} } \right)^{\nu_i} \exp \left[-\dfrac{{\cal R}}{{\cal R}_{\rm c} } \right] ~,
\end{equation}
for the $i$-th composition. In this work, the diffusion-reacceleration model is applied for the three propagation frameworks. The spatial distribution of CRs is obtained by numerically solving the diffusion equation with the DRAGON package. The diffuse gamma-ray distribution around the Galactic disk is calculated by the GALPROP package.

\subsection{CR energy spectra}

\begin{figure}
\centering
\includegraphics[height=6.cm, angle=0]{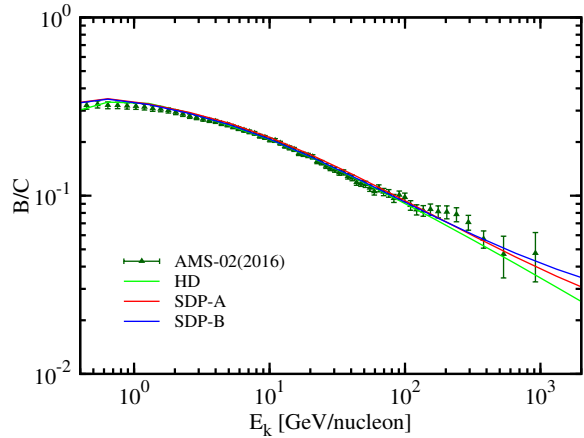}
\caption{
Calculation of B/C ratio to obtain the propagation parameters in HD, SDP-A and SDP-B models, with B/C data taken from the AMS-02 measurement \citep{2016PhRvL.117w1102A}.
}
\label{fig:bcratio}
\end{figure}

Before calculating the diffuse gamma-ray distribution from the $\pi^0$-decay, the CR spatial distribution in the Galactic halo have to be evaluated by certain propagation and injection parameters. For this purpose, the local observations of CR energy spectrum have to be fitted first, in order to obtain the propagation and injection parameters. In the HD propagation, the essential propagation parameters includes $D_0, \delta_0$ and $L$. As for the SDP models, three additional parameters have to be involved, i.e. $N_m, \xi$ and $n$. Fig. \ref{fig:bcratio} shows the fitting to the latest B/C ratio published by AMS-02 experiment \citep{2016PhRvL.117w1102A}, and the corresponding propagation parameters are listed in Tab. \ref{tab:para_diffu}. Compared with the conventional propagation, the SDP models anticipates a flattening for the B/C ratio above hundreds of GeV.

\begin{table*}
	\begin{center}
		\begin{tabular}{|c|ccccccc|}
			\hline
			Model & $D_{0}$ & ~~~$\delta_{0}$~~~ & ~~~$N_{m}$~~~  & ~~~$\xi$~~~ & ~~~$n$~~~~~ & ~~~~~$v_{A}$~~~ & ~~~$z_{h}$~~~ \\
			\hline
			& $[{\rm cm}^2\cdot {\rm s}^{-1}]$ &  &  &  &  & $[\rm km\cdot {\rm s}^{-1}]$ & $[\rm kpc]$ \\
			\hline
			HD & $4.55\times 10^{28}$    & 0.44 &  &  &  & 22 & 5 \\	
			SDP-A & $5.34\times 10^{28}$ & 0.58  &  0.51  & 0.1 & 4.0 & 6 & 5 \\
			SDP-B & $5.34\times 10^{28}$ & 0.58 & 0.4  & 0.1 & 3.5 & 6 & 5\\
			\hline
		\end{tabular}\\
	\end{center}
	\caption{Diffusion parameters of different propagation models.}
	\label{tab:para_diffu}
\end{table*}

\begin{figure*}
\centering
\includegraphics[height=8.cm, angle=0]{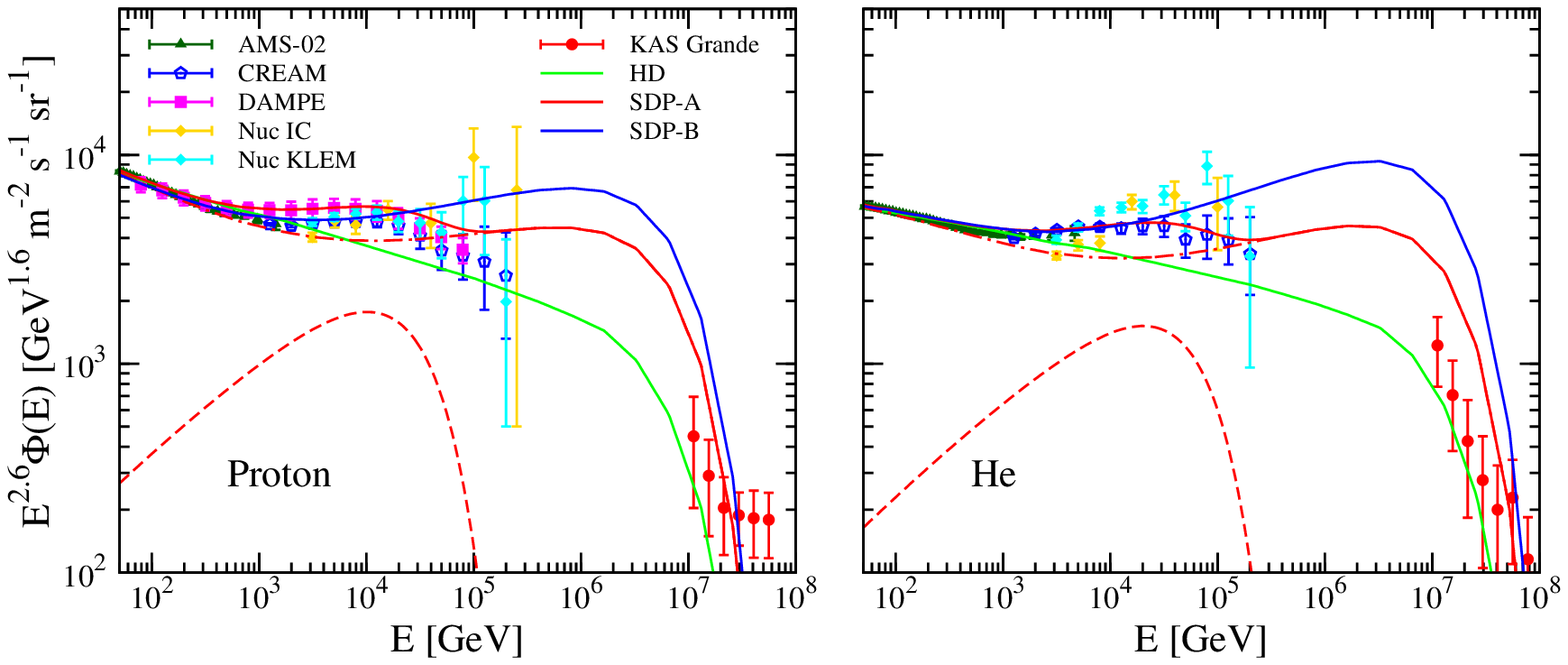}
\caption{
Calculated proton and helium energy spectra in propagation models of HD, SDP-A and SDP-B. The proton and helium data are taken from AMS-02 \citep{2015PhRvL.114q1103A,2017PhRvL.119y1101A}, CREAM-II \citep{2017ApJ...839....5Y}, NUCLEON \citep{2017JCAP...07..020A}, DAMPE \citep{2019SciA....5.3793A} and KASCADE \citep{2013APh....47...54A}.
}
\label{fig:phe1}
\end{figure*}

Given the propagation parameters, we further fit the local CR energy spectra to obtain the source's injection parameters. The calculated proton and helium spectra are illustrated in Fig. \ref{fig:phe1}, both of which are the major elements of CRs less than $10$ PeV. The injection parameters are listed in Tab. \ref{tab:para_inj}. The green lines are the propagated proton and helium spectra in the HD model. For the HD model, we do not taken into account the local source, since the nuclei excess between $200$ GV and tens of TeV does not affect the sub-PeV gamma-ray photon productivity. Compared with the observations less than $100$ TeV, both proton and helium spectra measured by the KASCADE experiment indicate a visible steepening. In this work, the cutoff rigidity of different elements is assumed to have $Z$-dependence. To fit KASCADE observations, the cutoff rigidity of proton and helium are set to $7$ PV for all propagation models.


\begin{table*}
	\begin{center}
		\begin{tabular}{|c|c|ccc|ccc|}
			\hline
			& & \multicolumn{3}{c|}{Background} & \multicolumn{3}{c|}{Local source} \\
			\hline
			Model & Element & Normalization$^\dagger$ & ~~~$\nu$~~~  & ~~~$\mathcal R_{c}$~~~ & ~~~$q_0$~~~~~ & ~~~~~$\alpha$~~~ & ~~~${\cal R}'_c$~~~ \\
			\hline
			& & $[({\rm m}^2\cdot {\rm sr}\cdot {\rm s}\cdot {\rm GeV})^{-1}]$ & & [PV] & [GeV$^{-1}$] & &  [TV] \\
			\hline
			HD & P   & $3.56\times 10^{-2}$    & 2.32   &  7  &  &  &  \\
			& He & $2.29\times 10^{-3}$   & 2.28     &  7  &  &  &   \\
			SDP-A & P   & $3.55\times 10^{-2}$   & 2.34    &  7  & $2.72\times 10^{52}$ & 2.1 & 28 \\
			& He & $2.28\times 10^{-3}$   & 2.28   &  7  & $2.53\times 10^{52}$ & 2.1  &   28  \\
			SDP-B & P   & $3.6\times 10^{-2}$   & 2.34    &  7  &  &  &  \\
			& He  & $2.4\times 10^{-3}$   & 2.26    &  7  &  &  &   \\
			\hline
		\end{tabular}\\
		$^\dagger${The normalization is set at kinetic energy per nucleon $E_{k} = 100$ GeV/n.}
	\end{center}
	\caption{Injection parameters of the background and local source in different models.}
	\label{tab:para_inj}
\end{table*}

The SDP-A model is marked as the red lines, in which the dash-dot, dash and solid lines denote the fluxes from background sources, local SNR and sum of them respectively. To reproduce the softening at $20$ TeV \citep{2017ApJ...839....5Y, 2019SciA....5.3793A}, the cutoff rigidities of local CRs are set to $28$ TV. It has been shown that this break is relevant to the peak in the anisotropy amplitude at $\sim 10$ TeV and the flip of anisotropy phase at $\sim 100$ TeV. In the SDP model, energetic CRs principally propagate within the IH. Therefore compared with HD, the background spectra (i.e. red dash-dot lines) gradually harden above $10^4$ GeV, so that the calculated fluxes above $10^5$ GeV are higher than HD model. The corresponding sub-PeV diffuse gamma-ray flux is expected to be higher in the SDP model.


The blue lines are the calculated proton and helium fluxes in SDP-B model. To interpret the spectral hardenging above $200$ GV, the parameter $N_m$ is $0.4$, which is smaller than $0.51$ of SDP-A model, so the propagated spectrum and corresponding B/C ratio are harder than those of SDP-A model. The cutoff rigidity is likewise set to $7$ PV in order to fit the KASCADE data. But the observed softening at around tens of TeV could not be explained, and both proton and helium fluxes inevitably exceed those of HD and SDP-A models above $100$ TeV. As we will show below, these extra fluxes are important to account for the observed sub-PeV diffuse gamma rays.

%

\subsection{Diffuse gamma-ray emission}

\begin{figure}
\centering
\includegraphics[height=6.cm, angle=0]{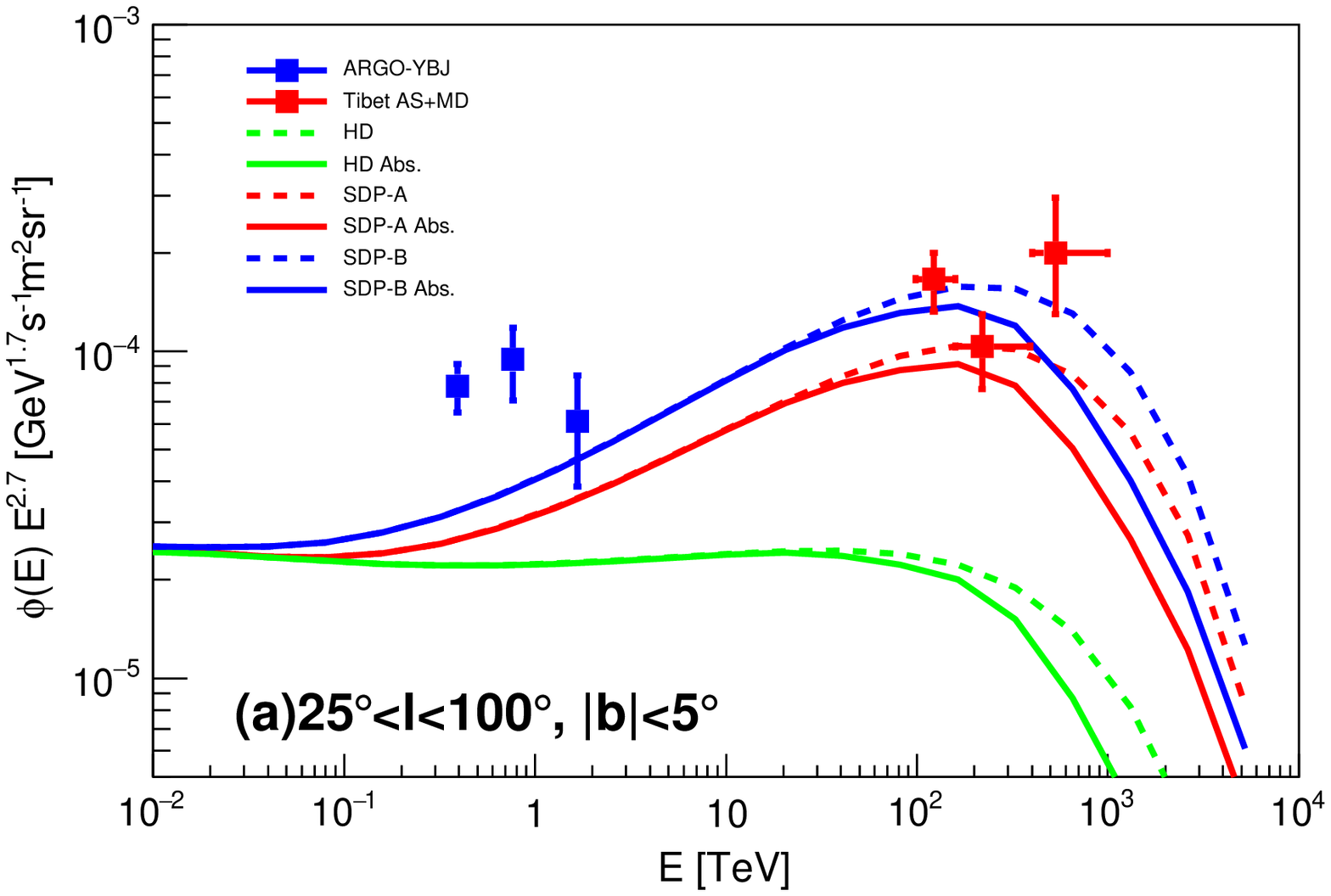}
\includegraphics[height=6.cm, angle=0]{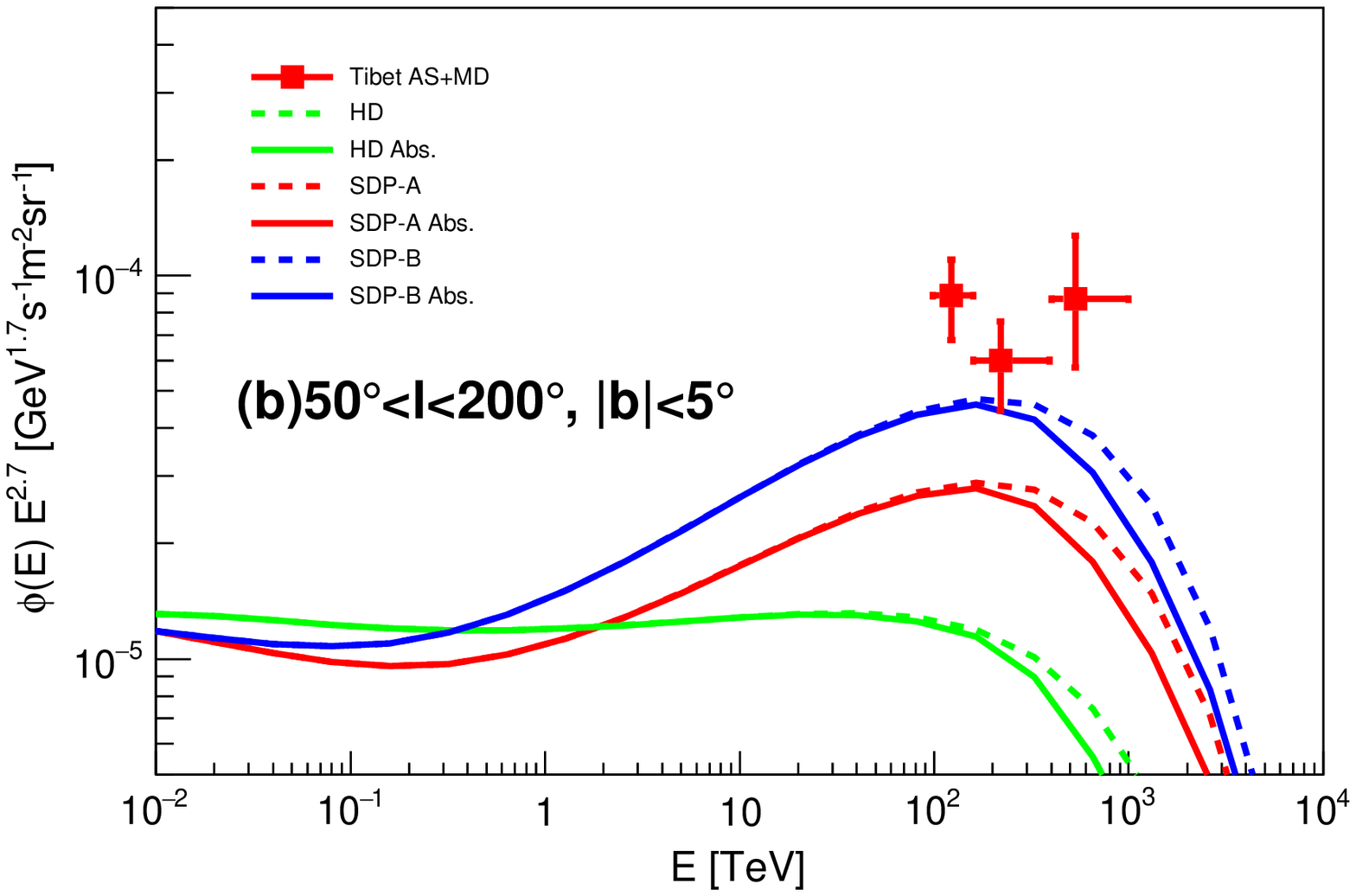}
\caption{
Calculated diffuse gamma-ray spectra in three propagation models. The gamma ray data are taken from ARGO-YBJ \citep{2015ApJ...806...20B} and Tibet AS+MD \citep{PhysRevLett.126.141101} experiments.
}
\label{fig:diffu_ga}
\end{figure}

After reproducing the local CR observations, we further calculate the spatial distribution of CRs and the corresponding diffuse gamma-ray spectra. Above $\sim 20$ TeV, the gamma-ray photons would have interactions with low-energy Galactic interstellar radiation field (ISRF) and the observed gamma-ray flux is expected to be strongly suppressed \citep{2006A&A...449..641Z, 2006ApJ...640L.155M}. Fig. \ref{fig:diffu_ga} shows the diffuse gamma-ray spectra calculated by the three propagation models. The solid and dash lines are the diffuse gamma-ray spectra with/without gamma-ray attenuation from pair production respectively. As illustrated in the figure, above $100$ TeV, the gamma-ray fluxes show observable attenuation. 

In HD model, the calculated diffuse gamma-ray fluxes are well below the observations. Compared with HD, the gamma-ray spectra above $1$ TeV show noticeable bumps in both SDP models so that the flux has been greatly enhanced correspondingly. This results from the spectral hardening of CRs between $100$ TeV and $10$ PeV in the entire halo. Meanwhile compared with SDP-B model, the diffuse gamma-ray flux in SDP-A model is lower, which means the CR flux in the whole Galaxy is always lower than SDP-B model above $100$ TeV. This is because the excess of CR flux become softening at $\sim 20$ TeV in SDP-A model, and is attributed to a local effect. Especially close to the direction of anti-Galactic center, i.e. $50^\circ < l < 200^\circ$, the gamma-ray flux above $100$ TeV is significantly lower than Tibet observation. Furthermore, the flux at $1$ TeV is slightly lower than ARGO measurements.



Only SDP-B model well describe both ARGO-YBJ and Tibet AS+MD observations at $25^\circ < l < 100^\circ$ and $50^\circ < l < 200^\circ$. This is due to the enhanced CR flux from $10$ TeV to $10$ PeV. Meanwhile in SDP-B model, the hardening is regarded as the propagation effect, which enables the CR flux above $10$ PeV in the entire diffusive halo to be augmented overall. Close to the Galactic center, the propagated CR energy spectra is harder.

\subsection{all-particle spectrum}

\begin{figure}
\centering
\includegraphics[height=6.cm, angle=0]{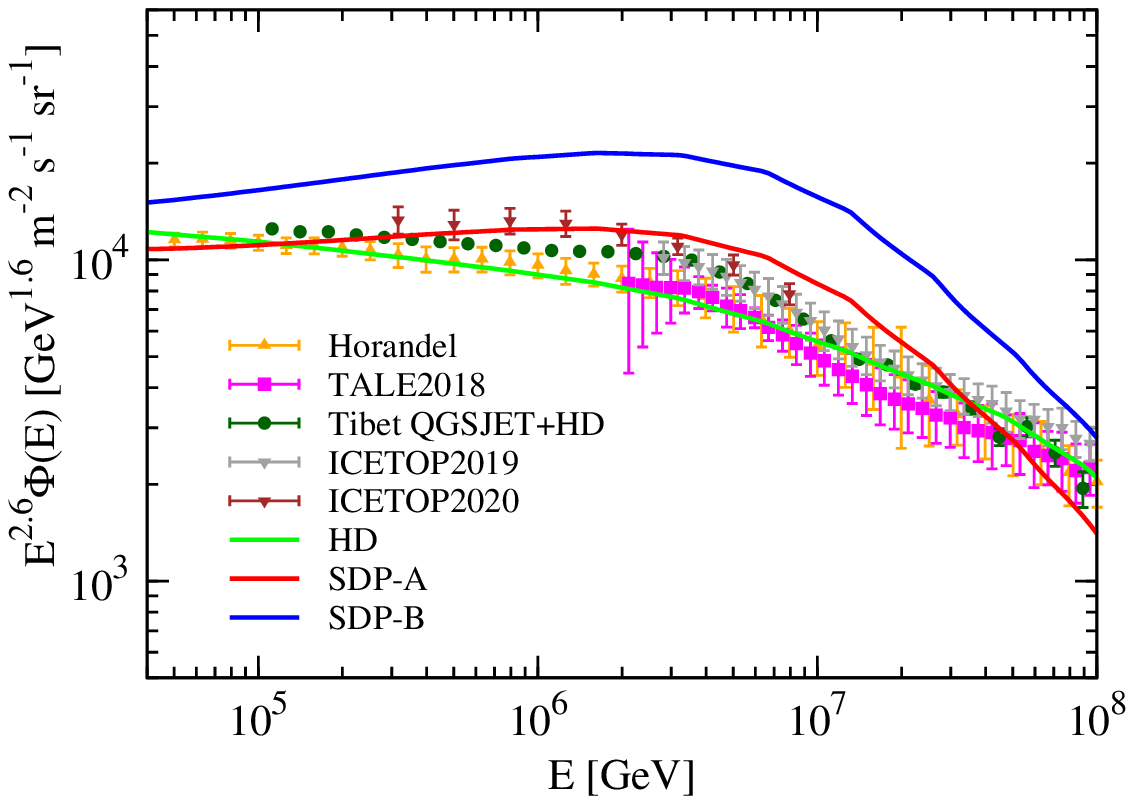}
\caption{
Calculated all-particle spectra in three propagation models. The all-particle data are taken from Horandel \citep{2003APh....19..193H}, TALE \citep{2018ApJ...865...74A}, IceTop \citep{2019PhRvD.100h2002A, 2020PhRvD.102l2001A} and Tibet \citep{2008ApJ...678.1165A}.
}
\label{fig:allparticle}
\end{figure}

The gamma-ray photons above hundreds of TeV are principally from the CRs at PeV energies, and an amount of ground-based observations have been performed at this energy range with high precision. In Fig. \ref{fig:allparticle}, we compare the calculated all-particle spectra with observations. The heavier nuclei are assumed to have the same power index with helium and the corresponding fluxes accommodate with local observations. The cutoff rigidity of nuclei is $Z$-dependent in order to fit the knee region. It is worth noting that less than knee region, which is most relevant to the Tibet sub-PeV gamma-ray observation, the major composition of CR flux is proton and helium.


We could see the all-particle spectrum in HD model is well consistent with Horandel spectrum in the whole energy range. For other experimental observations, the fitted all-particle spectrum are very close to them. For SDP-A model, there is a bit of excess above knee region, even for the latest ICETOP measurement in order to meet the gamma-ray measurement at hundreds of TeV. More severely, the all-particle flux in SDP-B model dramatically exceeds the observations from $100$ TeV to $100$ PeV, where iron is dominated. Especially, from $100$ TeV to knee region, where proton and helium fluxes are dominated, the all-particle flux is already higher than the observations.

As can be seen, the propagation origin of sub-PeV diffuse gamma-ray flux has a tension with local CR observations more or less. When local CR energy spectra are described, the calculated gamma-ray flux is inadequate for the Tibet observations. But if the sub-PeV gamma-ray flux is accounted for, the required CR flux surely exceeds the local CR energy spectra. This is obvious at $50^\circ < l < 200^\circ$, which means an unexpectedly large CR flux at anti-Galactic direction. In addition, the anisotropy observations are difficult to explain in SDP-B model, in which the anisotropy always points to the Galactic center, opposed to the observation below $100$ TeV.


\subsection{Diffuse neutrino spectrum}

Based on the sub-PeV gamma-ray observation, we could reckon the Galactic contribution below PeV. On average, the p-p collision produces nearly one-third neutral pions and two-thirds charged pions. Each neutral pion decays into a pair of gamma rays, whereas each charged pion decays into two muon neutrinos and one electron neutrino (here we do not distinguish between neutrinos and anti-neutrinos). The initial neutrino flavor ratio is approximately $\nu_e:\nu_\mu:\nu_\tau = 1:2:0$ from charged pion decay. After travelling, The flavor ratio is transformed to $\nu_e:\nu_\mu:\nu_\tau = 1:1:1$ due to the vacuum neutrino oscillation. The typical neutrino energy from charged pion decay is approximately half of the gamma ray photon from neutral pion decay. The gamma-ray spectrum at source is $\dif \Phi_\gamma /\dif E_\gamma  = \phi_\gamma E_\gamma^{-\Gamma}$, the resulting neutrino spectrum is shifted relative to the gamma ray spectrum \citep{2006PhRvD..74f3007K}, i.e.
\begin{align}
\dfrac{\dif \Phi_\nu}{\dif E_\nu}  =  \phi_\nu E_\nu^{-\Gamma} =  \left(\dfrac{1}{2} \right)^{\Gamma-1} \phi_\gamma E_\gamma^{-\Gamma} ~.
\end{align}

\begin{figure}
\centering
\includegraphics[height=6.cm, angle=0]{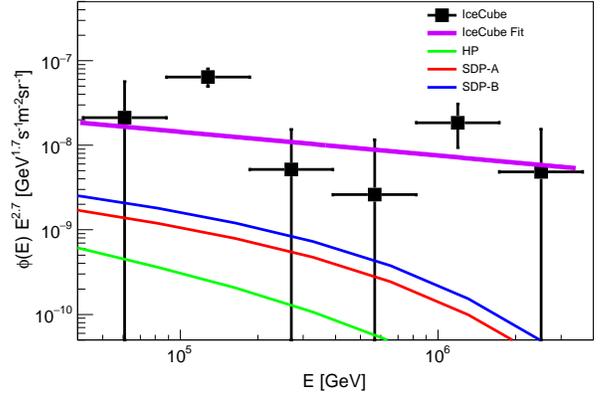}
\caption{
Diffuse neutrino flux calculated by the three propagation models. The data are taken from the ICECUBE $7.5$ years' observation \citep{2020arXiv201103545A}. The violet line is the power-law fitting to the data, with normalization $\Phi = 6.37 \times 10^{18}$ GeV$^{-1}$ cm$^{-2}$ s$^{-1}$ sr$^{-1}$ at $100$ TeV and power index $\gamma = -2.87$.
}
\label{fig:neu}
\end{figure}

In Fig. \ref{fig:neu}, we evaluate the diffuse neutrino flux in three propagation models. Compared with the fitting of ICECUBE observations, which is $\sim 2.87$, the spectrum of Galactic diffuse neutrinos is softer above hundreds of TeV. In the SDP-B model, the generated neutrino flux is largest compare with the other two models. Even in this case, the Galactic diffuse neutrinos generated during CR propagation, only takes up to at most $15 \%$ of latest observations. 

\section{Conclusion}

The argument that the CRs below knee region are accelerated by the Galactic sources, has long been the lack of clear evidence. Most recently, the Tibet AS$\gamma$ experiment reported their findings of the diffuse gamma ray photons between $100$ TeV and $1$ PeV, nearby the Galactic disk. These sub-PeV photons are likely the $\pi^0$-decay products from the hadronic interactions between the CR nuclei and interstellar materials. The Galactic origin of CRs with energy up to $10$ PeV have a definite conclusion along with observations of Galactic center, SNR G106.3+2.7 as well as Cyngnus Cocoon.



In this work, we study the propagation origin of these sub-PeV diffuse gamma rays by considering three competing propagation models. One is homogeneous diffusion and the other two are SDP models. Compared with the homogeneous diffusion, the SDP could generate more gamma rays above $10$ TeV around the disk due to the flattenning propagated CR spectra above tens of TeV. Nonetheless, to meet the observed sub-PeV gamma-ray flux at $25^\circ < l < 100^\circ$, the local CR flux in the SDP plus local source model has to slightly exceed the observation of the all-particle spectrum at around $10$ PeV. And the calculated gamma rays are still inadequate for the measurements of $1$ TeV at $25^\circ < l < 100^\circ$ and above $100$ TeV at $50^\circ < l < 200^\circ$. As for the SDP-B model, all above gamma-ray observations could be accounted for. However such a model meets up with severe challenges. The local CR flux would be greatly enhanced, which far exceeds the all-particle measurements. Meanwhile the softening of CR nuclei at $\sim 20$ TeV could not be reproduced, much less the observations of large-scale anisotropies.

%

In the current propagation models, there is a tension between sub-PeV gamma ray observation and local CR measurements. One possibilities is that the propagated spectrum is harder close to the Galactic center, whereas the propagation around the solar system is still unchanged, as indicated by the analysis of \cite{2016PhRvD..93l3007Y} for the radial distribution of diffuse $\gamma$-rays. Therefore the observed sub-PeV gamma rays and local CRs could be simultaneously satisfied by the SDP model. Meanwhile these sub-PeV gamma-ray photons from the Galactic disk may not entirely come from the CR propagation. Most of CR sources and interstellar medium are located around the Galactic disk. There may be some of gamma-ray photons from CR sources, despite that the events within $0.5^\circ$ from the known TeV sources have been subtracted. This has also be confirmed by the Tibet observation in fact. Above $398$ TeV, $4$ events at $50^\circ < l < 200^\circ$ are detected less than $4^\circ$ from the center of the Cygnus cocoon, which is just proved to be a PeV source by HAWC experiment. We hope more extensive research of diffuse gamma rays and the observation of single CR composition between $100$ TeV and $10$ PeV with high precision could testify our conclusion.

We evaluate the corresponding neutrino flux from the Galactic comic ray propagation. We find that even if the observed sub-PeV diffuse gamma rays could be accounted for, the Galactic halo could contribute $\sim 15\%$ of observed neutrino flux.




\section*{Acknowledgements}
This work is supported by the National Key Research and Development Program of China (No. 2016YFA0400200), the National Natural Science Foundation of China (Nos. U1738209, 11875264, 11635011, U2031110).

Software: GALPROP (\cite{1998ApJ...509..212S, 2000ApJ...537..763S}) available at https://galprop.stanford.edu.

DRAGON (\cite{2008JCAP...10..018E, 2017JCAP...02..015E}) available at https://github.com/cosmicrays.

\bibliographystyle{apj}
\bibliography{ref}

\end{document}